\documentstyle[sprocl,amssymb,epsfig]{article}

\input{psfig}

\def\Journal#1#2#3#4{{#1} {\bf #2}, #3 (#4)}

\def\NPB{{\em Nucl. Phys.} B}
\def\PLB{{\em Phys. Lett.}  B}

\def\PRD{{\em Phys. Rev.} D}

\def\be{\begin{equation}}
\def\ee{\end{equation}}
\def\bea{\begin{eqnarray}}
\def\eea{\end{eqnarray}}

\begin{document}
\begin{flushright}  
\large PITHA 98/20\\
\large hep-ph/9806481
\end{flushright}
\vskip1cm

\pagestyle{empty}

\begin{center}

{\LARGE \bf Progress in the Understanding\\
 of Dijet Production \\
in Deep-Inelastic Scattering\\}

\vskip1.2cm
 
{\large M. Wobisch}\\ \vskip5mm
III. Physikalisches Institut, RWTH Aachen \\
D-52056 Aachen, Germany
\end{center}
\vskip2cm

\begin{center}
\bf Abstract
\end{center}
Recent results on dijet production
in deep-inelastic scattering from the H1 experiment
at the ep-collider HERA are presented.
Internal jet structure has been studied in terms of 
jet shapes and subjet multiplicities in the Breit frame.
Both observables are seen to be well described by QCD models.
We observe a broadening of the jets towards the proton direction
and at lower transverse jet energies.
Dijet rates and dijet cross sections have been measured over a wide 
range of four momentum transfers ($5 < Q^2 < 5000\;\mbox{GeV}^2$) 
and transverse jet energies 
($25 < E^2_{t,\mbox{\tiny Breit}} \lesssim 1200\;\mbox{GeV}^2$)
with different jet algorithms.
Perturbative QCD calculations in next-to-leading order in the 
strong coupling constant give a good description of the data.

\vskip20mm
{\small Talk given on behalf of the H1 collaboration
at the 6th International Workshop on Deep Inelastic Scattering and QCD 
(DIS 98), Brussels, Belgium, 4-8 April 1998.}

\newpage

\pagestyle{plain}
\pagenumbering{arabic}

\title{Progress in the Understanding of Dijet Production in DIS}

\author{M. Wobisch}

\address{III. Physikalisches Institut, RWTH Aachen,\\
D-52056 Aachen, Germany,\\
E-mail: Markus.Wobisch@desy.de} 


\maketitle\abstracts{Recent results on dijet production
in deep-inelastic scattering from the H1 experiment
at the ep-collider HERA are presented.
Internal jet structure has been studied in terms of 
jet shapes and subjet multiplicities in the Breit frame.
Both observables are seen to be well described by QCD models.
We observe a broadening of the jets towards the proton direction
and at lower transverse jet energies.
Dijet rates and dijet cross sections have been measured over a wide 
range of four momentum transfers ($5 < Q^2 < 5000\;\mbox{GeV}^2$) 
and transverse jet energies 
($25 < E^2_{t,\mbox{\tiny Breit}} \lesssim 1200\;\mbox{GeV}^2$)
with different jet algorithms.
Perturbative QCD calculations in next-to-leading order in the 
strong coupling constant give a good description of the data.}

\section{Introduction}
The production of jets with high transverse energies in the Breit
frame in deep-inelastic scattering is directly sensitive to the dynamics of the
strong interaction. 
Perturbative calculations in next-to-leading 
order in the strong coupling constant are expected to be able
to describe this process.
In this contribution we present various measurements of dijet production
in deep-inelastic scattering.
We report on extensions of measurements of dijet rates 
shown at the previous DIS workshop \cite{mwdis97} in a 
slightly modified phase space.
We have also performed new measurements of dijet cross sections 
at higher transverse jet energies in the Breit frame with
different jet algorithms.
Jet shapes and subjet multiplicities in dijet production in the Breit frame
are used to study the internal structure of jets.
All measurements presented here have been corrected for detector effects
(corresponding to the level of stable hadrons).

\section{Internal Jet Structure}
The understanding of the internal structure of jets is an important 
prerequisite for the understanding of the production rates of jets.
Furthermore it also allows a deeper insight
into the mechanism of how hard partons evolve into jets of hadrons.
In the present analysis we have studied two observables using cone
and $k_t$ jet algorithms: subjet multiplicities and jet shapes.

\begin{figure}[t]
\epsfig{file=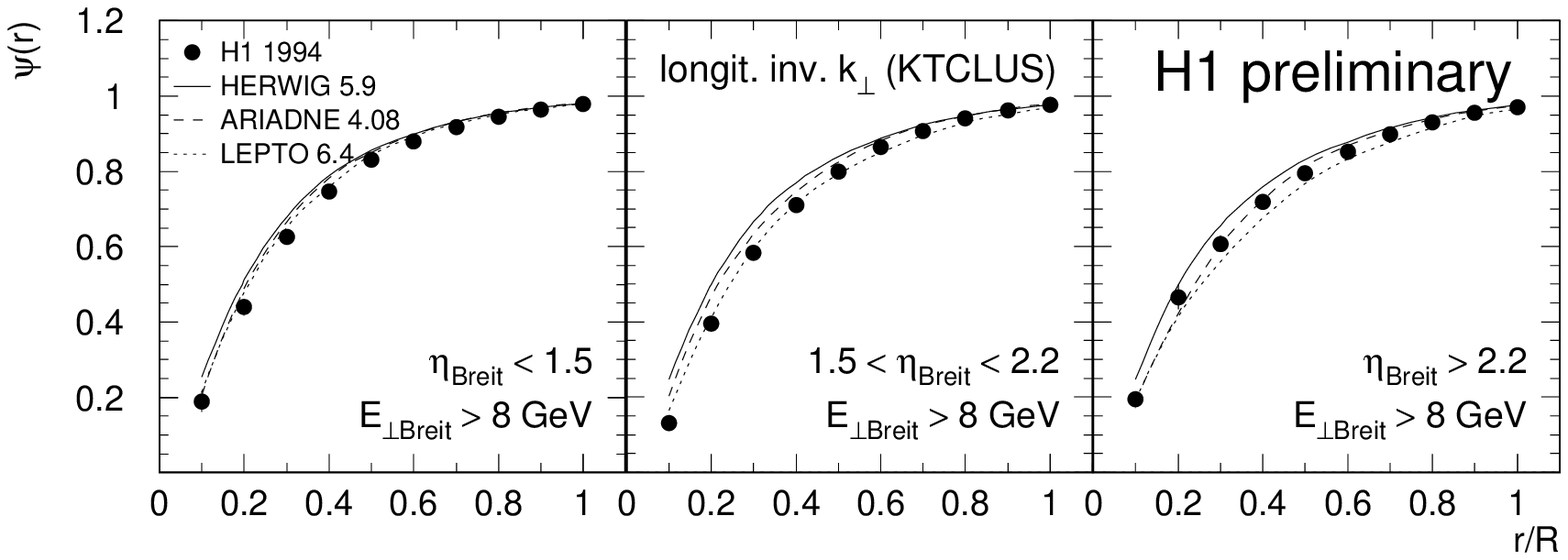,height=1.84in}\vskip-1mm 
\epsfig{file=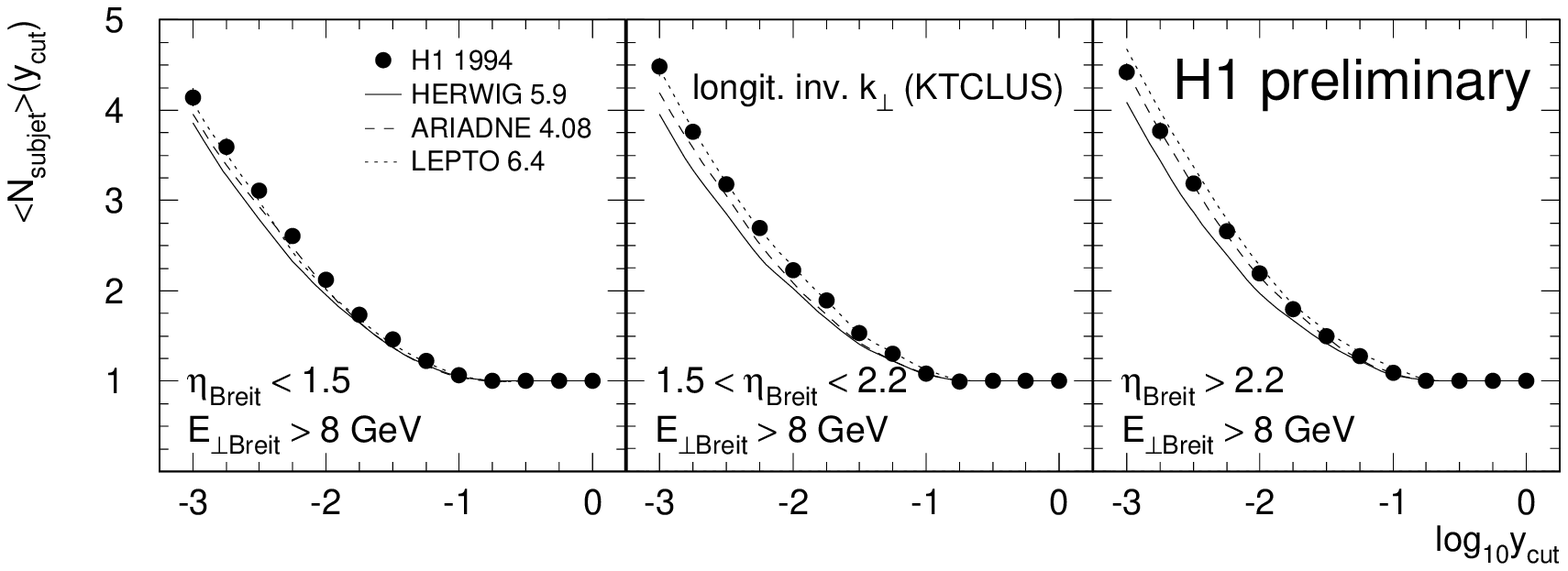,height=1.84in}\vskip-0mm 
\caption{Observables of internal jet structure for the $k_t$ cluster 
algorithm as a function of the jet pseudorapidity in the Breit frame 
for jets with transverse energies of 
$E_{t,\mbox{\scriptsize jet, Breit}} > 8\,\mbox{GeV}$.
Jet shapes (top) as a function of the radius $r$ relative to the jet 
radius $R$ and subjet multiplicities (bottom) as a function of the 
resolution parameter $y_{\mbox{\scriptsize cut}}$ (as explained in the text)
are compared to the prediction of QCD models. 
Positive pseudorapidities are towards the proton direction.
\label{fig:jetstr}}
\end{figure}

{\bf The jet shape} $\psi(r)$
is defined as the average fractional transverse energy of the jet
inside a subcone of radius $r$ concentric around the jet axis.
Only particles assigned to the jet are considered \cite{MHS}. 
The measurements have been performed using a $k_t$ \cite{ktincl}
and a cone \cite{snowmass} jet algorithm.

{\bf Subjet multiplicities} are a natural way of studying internal 
jet structure for clustering algorithms.
The clustering procedure is repeated for all particles assigned to a 
given jet. 
The clustering is stopped when the distances $d_{ij}$ between 
all particles $i,j$ are above some cutoff $y_{\mbox{\scriptsize cut}}$
$$
d_{ij}  \; =  \; \min (E_{t,i}^2 , E_{t,j}^2) \, / \, 
E_{t,\mbox{\scriptsize jet}}^2  \,
\cdot \, ( \Delta \eta_{ij}^2 + \Delta \varphi_{ij}^2 )
 \; >  \; y_{\mbox{\scriptsize cut}} \; .
$$
The remaining (pseudo-)particles are called subjets.
The observable studied in this analysis is the average number
of subjets at a given value of $y_{\mbox{\scriptsize cut}}$ 
in the range $10^{-3} \le y_{\mbox{\scriptsize cut}} \le 1$
for the $k_t$ algorithm.

The measurements have been performed in a region
of four momentum transfers $10 < Q^2 \lesssim 120\,\mbox{GeV}^2$ 
and values of the inelasticity variable of $0.15 < y < 0.6$.
The jet algorithms are applied in the Breit frame.
Events with at least two jets with transverse energies of
$E_{t,\mbox{\scriptsize jet,Breit}} > 5\,\mbox{GeV}$ inside
the pseudorapidity region in the laboratory frame 
$-1 < \eta_{\mbox{\scriptsize lab}} < 2$ are selected.

Both the jet shapes and the average number of subjets are presented
as a function of the transverse energy 
$E_{t,\mbox{\scriptsize jet,Breit}}$ and the pseudorapidity 
$\eta_{\mbox{\scriptsize jet,Breit}}$ of the jets in the Breit 
frame\footnote{Positive pseudorapidities are along the positive
z-axis which is defined as the direction of the incoming proton in
both, the Breit and the laboratory frame.}.
The results for the $k_t$ and for the cone algorithm are well described 
by QCD models (in Fig.\ \ref{fig:jetstr} we show some results for the 
$k_t$ algorithm; see also \cite{tcdis98}).
We observe a dependence of the jet broadness and the average number of subjets
on the transverse energy and the pseudorapidity of the jets in the Breit frame.
With increasing transverse jet energies and decreasing pseudorapidities
the jets are more collimated.
Furthermore the jets defined by the $k_t$ algorithm turn out to be 
narrower than jets defined by the cone algorithm.

\section{Dijet Rates}
At the previous DIS workshop we have presented measurements of dijet 
rates \cite{mwdis97} (i.e.\ the fraction of dijet events in all DIS events)
for a cone algorithm at four momentum 
transfers $5 < Q^2 < 100\,\mbox{GeV}^2$ and transverse jet energies
$E_{t,\mbox{\scriptsize jet,Breit}} > 5\,\mbox{GeV}$.
The predictions of next-to-leading order (NLO) 
calcula\-tions \cite{disent} were significantly lower
(up to factors of 0.5) than the data.

In the meantime it has been argued \cite{symmcut} that the phase space 
chosen in that analysis contained infrared sensitive regions
where fixed order calculations are not predictive.
This mainly concerns the selection cut on the transverse jet 
energies $E_{t,\mbox{\scriptsize min,Breit}} = 5\,\mbox{GeV}$ 
which is applied for both jets.
For this selection cut the cross section receives large contributions
from the threshold region where both jets have similar transverse energies
$E_{t,1} \simeq E_{t,2} \simeq E_{t,\mbox{\scriptsize min,Breit}}$.
The emission of a soft gluon can reduce the transverse energy of one 
of the jets such that it falls below the threshold.
In fixed order calculations this results in an incomplete cancelation 
between real and virtual corrections at the threshold region, 
thereby making them unpredictive.

The infrared sensitive regions can e.g.\ be avoided by asymmetric $E_t$ cuts
($E_{t,\mbox{\scriptsize min}} > 5\,\mbox{GeV}$ and
$E_{t,\mbox{\scriptsize max}} > 7\,\mbox{GeV}$)
or by applying an additional harder cut on the sum of the transverse 
jet energies ($E_{t,\mbox{\scriptsize min}} > 5\,\mbox{GeV}$ and
$\sum_{1,2} E_{t,i} > 13\,\mbox{GeV}$).
We have repeated the analysis for both of these scenarios, using 
the same selection cuts as in the previous analysis otherwise \cite{mwdis97}.

\begin{figure}[t]
\begin{center}
\epsfig{file=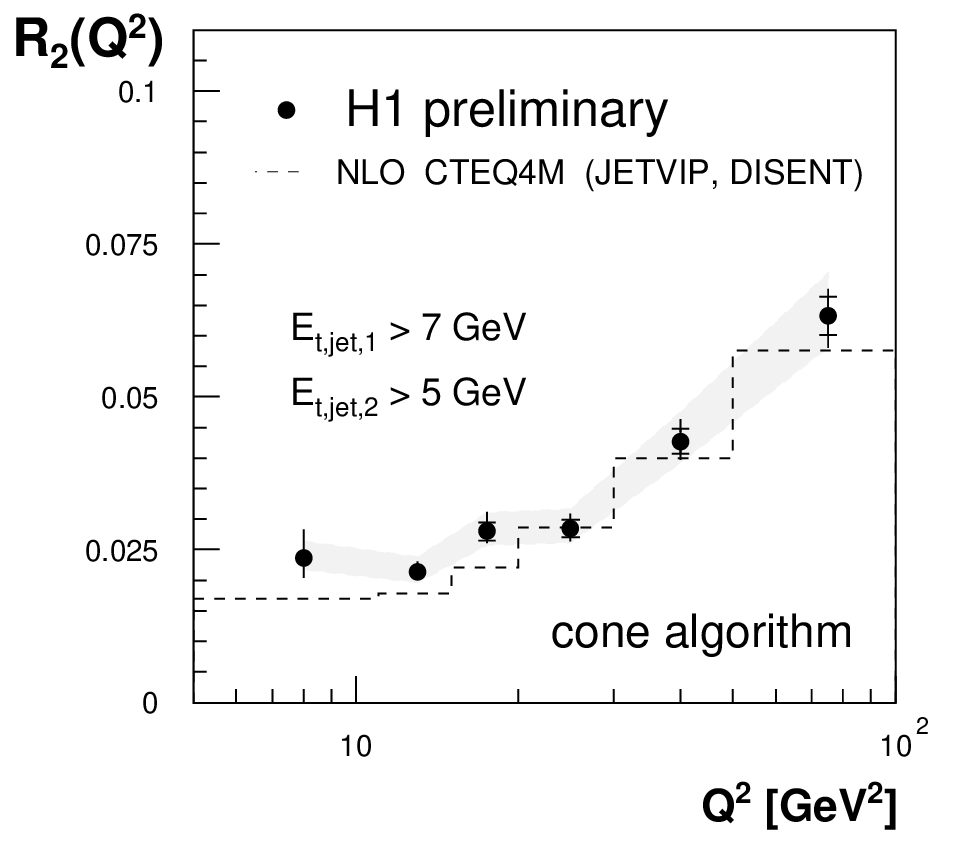,width=2.25in}\epsfig{file=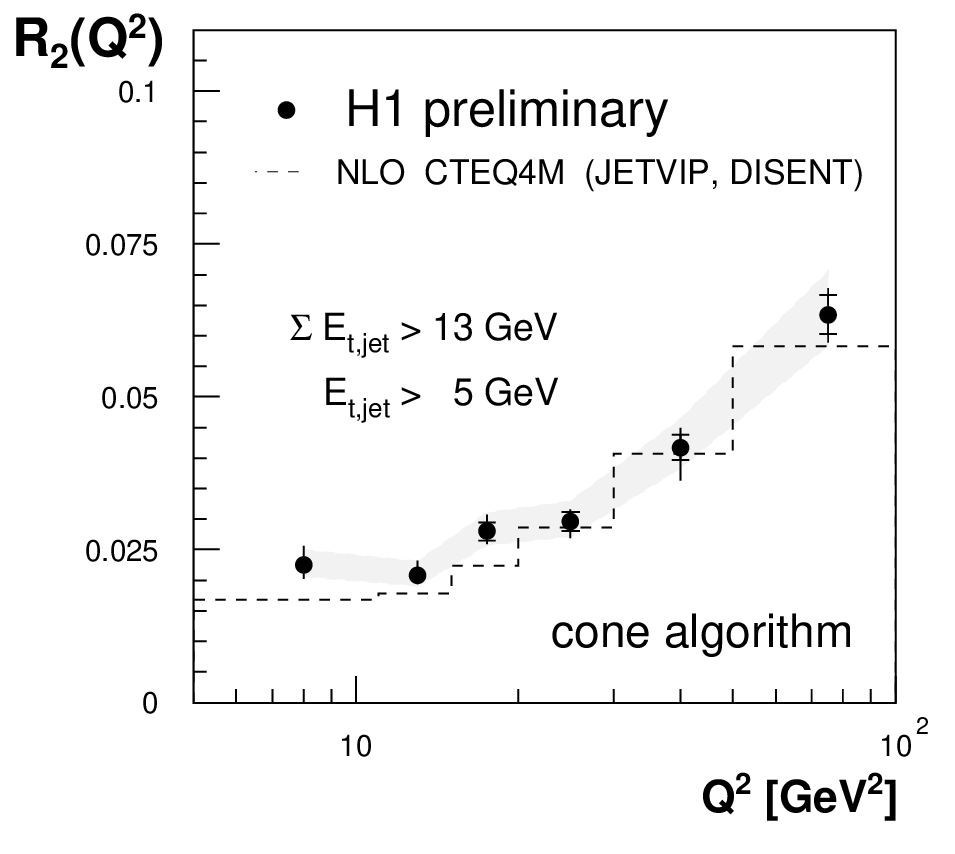,width=2.25in}\vskip-2mm
\end{center}
\caption{The $Q^2$ dependence of the dijet rates for transverse jet energy
cuts that avoid infrared sensitive phase space regions.  
The data are compared to NLO predictions with the renormalization
and the factorization scale set to 
$\mu^2_r = \mu^2_f = Q^2 + \langle E_t \rangle^2 $.
\label{fig:dijetrates}}
\end{figure}

The results are displayed in Fig.\ \ref{fig:dijetrates} where
the data are compared to NLO predictions\footnote{Hadronization corrections
are not included in the comparison. Their size has been estimated 
by QCD models to be in the order of 10\,\% (i.e. the partonic 
dijet cross section is 10\,\% higher as compared to the hadronic 
dijet cross section).} \cite{disent} \cite{bjoerndis98}.
It is noticeable that the data are equally well described by the NLO 
calculations in both cases, so that it does not seem to play a role
how in detail the infrared sensitive regions are avoided.
Further discussion on these data can be found in \cite{bjoerndis98}.

\section{Dijet Cross Sections}
The last analysis that we present here has been performed 
with different jet clustering algorithms:
the longitudinally invariant $k_t$ algorithm (originally proposed for
hadron collisions \cite{ktincl}), 
the $k_t$ algorithm that was originally proposed for DIS \cite{ktdis}, 
and the recently proposed Cambridge algorithm \cite{cambridge} 
(which we have modified for DIS to treat the proton remnant 
according to the prescription used in the $k_t$ algorithm for DIS).

The dijet analysis covers a kinematical range of four momentum transfers
$200 < Q^2 < 5000\,\mbox{GeV}^2$ (for the longitudinally invariant $k_t$
the range was extended down to $Q^2 = 10\,\mbox{GeV}^2$) and
$0.2 < y < 0.6$.
For the longitudinally invariant $k_t$ algorithm we required
$E_{t,\mbox{\scriptsize min, Breit}} > 5\,\mbox{GeV}$ and
$\sum_{1,2} E_{t,i} > 17\,\mbox{GeV}$,
while the $k_t$ for DIS and the Cambridge algorithm were used
with a reference scale of $100\,\mbox{GeV}^2$ and
$y_{\mbox{\scriptsize cut}} =1$
(these values were chosen to obtain dijet cross sections of similar
size for all algorithms).
Events with at least two jets in 
$-1 < \eta_{\mbox{\scriptsize lab}} < 2.5$ are selected.

\begin{figure}[t]
\begin{center}
\epsfig{file=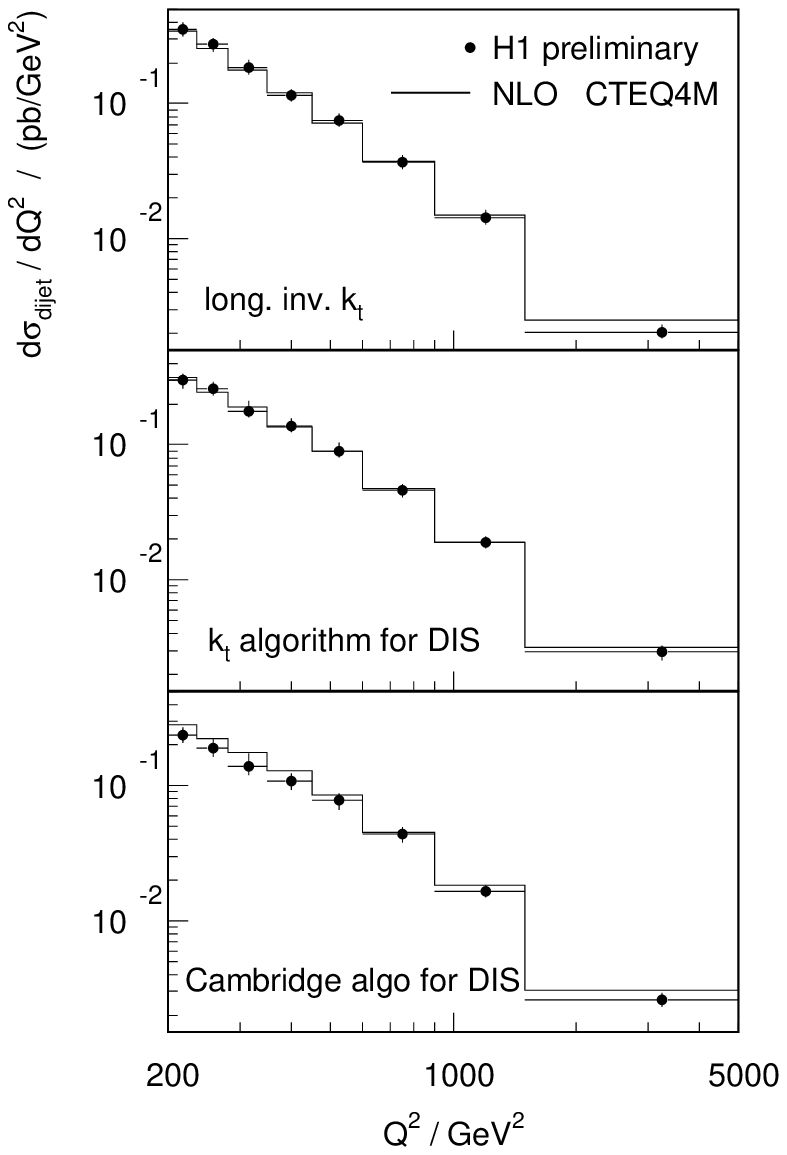,width=2.25in}\epsfig{file=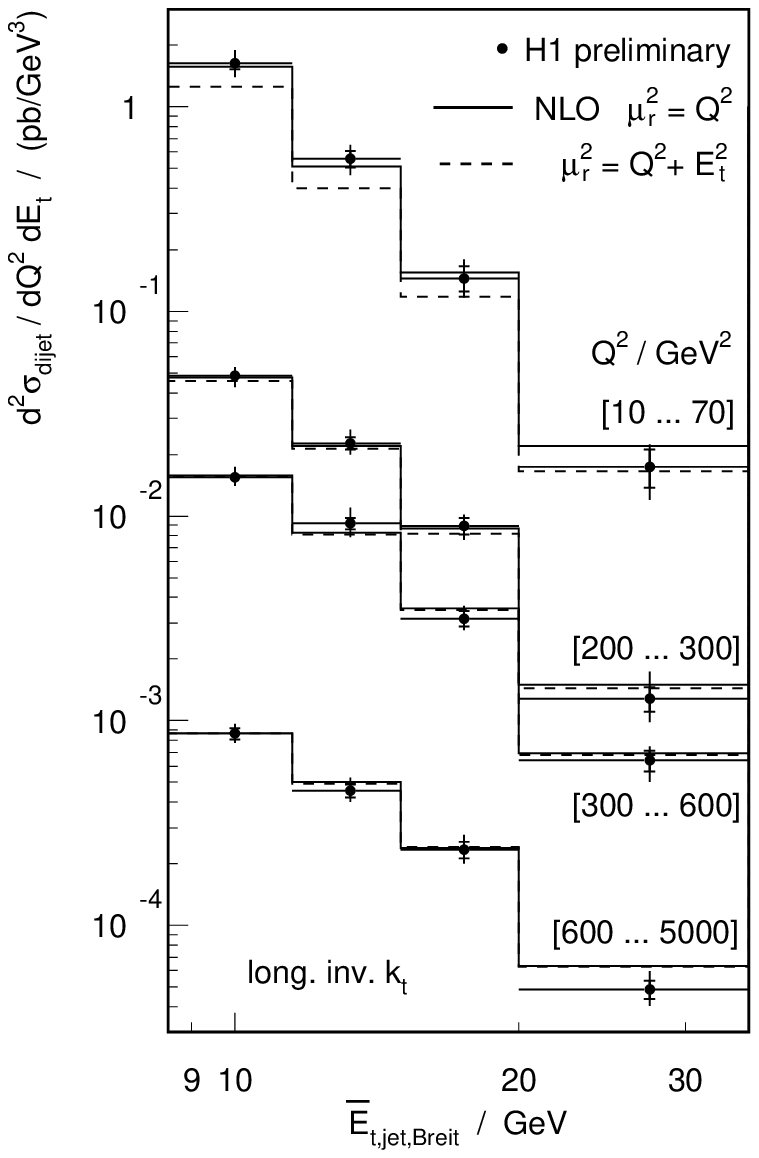,width=2.25in}
\end{center}
\caption{The $Q^2$ dependence of the dijet cross sections for 
different jet algorithms (left) and the average transverse jet energy 
distribution of the two jets with highest transverse energies 
for the longitudinally invariant $k_t$ algorithm (right)
compared to NLO predictions. \label{fig:dijetcr}}
\end{figure}

The results are presented in Fig.\ \ref{fig:dijetcr} where the data
are compared to the NLO predictions.
Hadronization corrections are not considered in the 
comparison\footnote{QCD models predict hadronization corrections of
7\,\% (longitudinally invariant $k_t$), and on average 15\,\% ($k_t$ for DIS)
and 18\,\%(Cambridge); the corrections for the latter algorithms show 
a stronger $Q^2$ dependence.}.
For all algorithms we observe good agreement between data and theory.
However, at lower $Q^2$ ($Q^2 \lesssim 100\,\mbox{GeV}^2$) the agreement 
depends strongly on the choice of the renormalization scale in the 
NLO calculation.
The choice of $\mu^2_r = Q^2$ gives a perfect description of the data 
(which is also seen in other dijet distributions not shown here).
Although a scale of $\mu^2_r = Q^2 + E^2_t$ is more reasonable 
(it gives a smooth interpolation between $\mu^2_r = E^2_t$ in the 
limit $Q^2 \rightarrow 0$ and $\mu^2_r = Q^2$ in the limit 
$E_t^2 \rightarrow 0$), this choice leads to NLO cross sections 
up to 30\,\% below the data.

\section{Summary}
Within the last year we have significantly improved our understanding 
of dijet production in deep-inelastic scattering.
Measurements of internal jet structure are very well described by QCD models.
We have learned that the applicability of perturbative calculations
in fixed order can be heavily restricted in badly chosen 
(i.e.\ infrared sensitive) regions of phase space.
Considering such limitations and avoiding infrared sensitive
phase space regions, we have shown that NLO calculations can give a very good 
description of our dijet data over a very large kinematical range of
four momentum transfers $10< Q^2 < 5000\,\mbox{GeV}^2$ and transverse jet
energies 
$25 < E^2_{t,\mbox{\scriptsize Breit}} \lesssim 1200\;\mbox{GeV}^2$,
for various jet definitions.


\section*{References}
\begin{thebibliography}{9}
\bibitem{mwdis97}M. Wobisch, Proceedings of the International Workshop 
on Deep Inelastic Scattering and QCD (DIS 97), Chicago (1997).
\bibitem{MHS}M.H. Seymour, \Journal{\NPB}{513}{269}{1998}.
\bibitem{ktincl} St.D. Ellis, D.E. Soper, \Journal{\PRD}{48}{3160}{1993};\\
S. Catani, Yu.L. Dokshitzer, M.H. Seymour, B.R. Webber, 
{\em Nucl. Phys.} B {\bf406}, 187 (1993).
\bibitem{snowmass}J.E. Huth et al.\, Proceedings of the 1990 DPF Summer Study
on High Energy Physics, Snowmass, Colorado, edited by E.L. Berger, 
World Scientific, Singapore, 134 (1992).
\bibitem{tcdis98} I. Bertram, et al., these proceedings.

\bibitem{disent} S. Catani, M.H. Seymour, \Journal{\NPB}{485}{291}{1997}.
\bibitem{symmcut} St. Frixione, G. Ridolfi, \Journal{\NPB}{507}{315}{1997}.
\bibitem{bjoerndis98} B. P\"otter, these proceedings.
\bibitem{cteq4} H.L. Lai, et al., \Journal{\PRD}{55}{1280}{1997}.

\bibitem{ktdis} S. Catani, Yu.L. Dokshitzer, B.R. Webber,
    \Journal{\PLB}{285}{291}{1992}.
\bibitem{cambridge} Yu.L. Dokshitzer, G.D. Leder, S. Moretti, B.R. Webber,
 JHEP 08(1997)001, CAVENDISH-HEP-97-06.

\end{thebibliography}
\end{document}